\documentclass[preprintnumbers,amsmath,18pt,amssymb,floatfix,superscriptaddress,nofootinbib]{article}

\def\mytitle#1{\setcounter{equation}{0}
\setcounter{footnote}{0}
\begin{flushleft}\Large\textbf{#1}\end{flushleft}
\vspace{0.25cm}}
\def\myname#1{\leftline{{\large #1}}\vspace{-0.13cm}}
\def\myplace#1#2{\small\begin{flushleft}\textit{#1}\\
\texttt{#2}\end{flushleft}}

\def\myclassification#1{\small\noindent
Keywords :
       #1\vspace{0.5cm}}
\usepackage{graphicx}
\usepackage{amssymb}
\usepackage{amsmath}
\usepackage{graphicx}
\usepackage{hyperref}
\usepackage{color}

\begin{document}
\mytitle{Cosmology with Distinct Functions $f$ of the Non-metricity Scalar $Q$ : A Dynamical System Approach}

\myname{$Promila~Biswas^{*}$\footnote{promilabiswas8@gmail.com}, $Subhajit~Pal^{**}$\footnote{subhajitpal968@gmail.com },$Ritabrata~
Biswas^{**}$\footnote{biswas.ritabrata@gmail.com} and $Satyajit~Pal^{**,~\dag}$\footnote{satyajitpal1995@gmail.com}}
\myplace{*Department of Mathematics, Jadavpur University, Kolkata-32, India\\ **Department of Mathematics, The University of Burdwan, Golapbag Academic Complex, Burdwan - 713104, Purba Burdwan, India.\\ \dag Department of Mathematics, Dr. Bhupendra Nath Dutta Smriti Mahavidyalaya, Hatgobindapur-713407, Purba Burdwan, India.} {}
\begin{abstract}
Symmetric teleparallel gravity is one among the general relativistic trinity which deals with the non-metricity scalar $Q$. In the Einstein Hilbert action, a function of $Q$ is chosen to be the main contributory part of the Lagrangian and a modified theory of gravity is constructed. In literature, different structures of the function of $Q$ are found which sustain several astrophysical observations like Big Bang nucleosynthesis, late-time cosmic acceleration etc. Autonomous systems for each such models with different $f(Q)$ structures are constructed. Corresponding fixed points and their stability properties are studied. For every case, stable, unstable and saddle-type fixed points are found to exist. These points on the phase portraits are cosmologically analyzed. It is tried to justify which way the corresponding state may lead if the initial state is perturbed. A comparative study of different models is represented.
\end{abstract}
\myclassification{Symmetric teleparallel gravity, $f(Q)$ gravity, fixed point analysis, cosmological equilibrium points.}\\
PACS No.:  04.50.Kd, 98.80.Jk, 02.40.Yy, 64.60.F .
\section{Introduction}
 
Newtonian gravity was geometrically represented in general relativity (GR) with the help of the curvature tensor \cite{Jim_nez_2018}
\begin{equation}
R^\alpha_{\beta \mu \gamma} \equiv 2 \partial_{[\mu}\Gamma^\alpha_{\nu]\beta}+2\Gamma^\alpha_{[\mu | \lambda |}\Gamma^\lambda_{\gamma]\beta}~~~~~~,
\end{equation}
where the $2^{nd}$ order Christoffel symbol $\Gamma_{\mu \nu}^{\beta} \equiv
\begin{Bmatrix}
\beta\\
\mu \nu
\end{Bmatrix}$ 
is constructed upon the mostly plus signed Lorentzian signatured metric $ g_{\mu \nu}$ as 
\begin{equation}\label{Chritoffel_2}
\begin{Bmatrix}
\beta\\
\mu \nu
\end{Bmatrix}
=\frac{1}{2}g^{\beta \sigma} \bigg(\frac{\partial g_{\sigma \lambda}}{\partial x^{\mu}} + \frac{\partial g_{\sigma \mu}}{\partial x^{\lambda}} - \frac{\partial g_{\mu \lambda}}{\partial x^{\sigma}}  \bigg)~~~~~.\end{equation}
This affine connection is metric compatible and torsionless. 

Just after this theory's proposition around $1916$, several observational success are attained. However, conceptual and technical difficulties are found. Concerned geometry can be simplified by vanishing $ R^\alpha_{\beta \mu \nu}$. This generates Weitzenb$\ddot{o}$ck \cite{Aldrovandi:2013wha} form. GR can be reformulated in a flat and torsion free spacetime with such improvements.

A vanishing Christoffel, i.e., $ \Gamma^\alpha_{\mu \nu}$ will result into a vierbein formulation of GR. This reinterpretation requires knowledge of the group $GL(4, \mathbb{R})$ \cite{ISHAM197198} and teleparallelized in the metric affine gauge theory \cite{PhysRevD.67.044016}.
The formalization requires the decomposition of general affine connection as 
$$\Gamma_{\mu \nu}^{\beta}=
\begin{Bmatrix}
\beta\\
\mu \nu
\end{Bmatrix}
 + K^\beta_{\mu \nu}+L^\beta_{\mu \nu}~~~~~~~,$$
 where the contortion $K_{\mu \nu}^\alpha$ comes as 
 $$K_{\mu \nu}^\alpha = \frac{1}{2}T^\alpha_{\mu \nu} - T_{(\mu^\alpha \nu)}~~~~~~~,$$
 with the torsion $T^\alpha_{\mu \nu}=2 \Gamma^\alpha_{[\mu \nu]}$.
The disformation tensor $L_{\mu \nu}^\beta$ is defined as
$$L_{\mu \nu}^\beta = \frac{1}{2}Q^\beta_{\mu \nu} - Q_{(\mu^\beta \nu)}~~~~~~.$$
As non-metricity implies $Q_\alpha \equiv Q_{\alpha \mu}^{\mu} $, these tensor fields satisfy important identities, as for example, Bianchi identity \cite{Ortin:2015hya},
\begin{equation}
R^\mu_{\alpha \beta \gamma}-\nabla_{[\alpha} T^\mu_{\beta \gamma]} + T^\gamma_{[\alpha \beta}T^\mu_{\gamma]\nu}=0~~~~~~~~~.
\end{equation}
The scalar action has the form 
\begin{equation}
{\cal S}=\int d^n x \sqrt{-g}f\left(g^{\mu \nu}, R^\alpha_{\beta \mu \nu}, T^\alpha_{\mu \nu}, Q^{\mu \nu}_\alpha \right) +{\cal S}_M~~~~~~~~,
\end{equation}
independent variables are the metric and the affine connection, ${\cal S}_M$ contains the matter field.

In the case of varying cosmological models, one is lucky enough to be armed with the associated dynamical systems analysis as a powerful weapon \cite{Rosa_2024}. By this method, without even solving complex differential equations, one may build an idea regarding the qualitative dynamics of the universe. Cosmological models, in this way of representation, are portrayed as phase spaces each point of which corresponds to a state of the concerned system. The evolution of the system is marked with trajectories that reveal the stability criteria as well \cite{Rosa_2024}.

To mark equilibrium states, first-order differentiation of the coordinate vector should vanish. From this condition, we may find cosmological equilibrium points. The nature of such fixed or equilibrium points tell us what impacts a small perturbation may lead to \cite{B_hmer_2016} long-term evolution of universe like the emergent universe, early inflation, dark energy dominance, bouncing universe etc can be speculated qualitatively.

In $f(Q)$ gravity, Einstein-Hilbert action is written as 
$${\cal S}=\int d^4x\sqrt{-g}\left[\frac{1}{2}f(Q)+{\cal L}_{matter}\right]~~~~,$$
which alters the cosmic dynamics as we generalize $f(Q)$ as $Q+F(Q)$ leading to modified Friedmann equations. For $f(Q)=Q+\alpha Q^2$ model, $\alpha$ is a constant, the nonlinear term dominates high non-metricity, i.e., the early universe which triggers inflation which mimics the effect of $R^2$ in the original Starobinsky model \cite{nojiri2024well}.

Using nonmetricity, inflation without a scalar field can be generated \cite{
capozziello2022slow}. On the other hand, it may offer graceful exit from inflation and can generate the necessary density perturbations\cite{crittenden1992graceful}. Inflation, quintessence or phantom can be mimiced by different choices of the function $f$ of nonmetricity scalar\cite{narawade2024modelling}. 

Now, we consider a dynamical system of the form:
$$
\frac{d\mathbf{\tilde{x}}}{dt} = \mathbf{\tilde{f}}(\mathbf{\tilde{x}})~~~~,
$$
where $\mathbf{\tilde{x}} = \left(x_1(t), x_2(t), \dots, x_n(t)\right)^T$ is a vector of state variables, and 
$$
\mathbf{\tilde{f}}(\mathbf{\tilde{x}}) = \left(f_1(\mathbf{\tilde{x}}), f_2(\mathbf{\tilde{x}}), \dots, f_n(\mathbf{\tilde{x}})\right)^T
$$
is a vector field describing the system's dynamics.

An equilibrium point \(\mathbf{\tilde{x}}_0\) is a point in the state space where
$$
\mathbf{\tilde{f} }(\mathbf{\tilde{x}}_0) = \tilde{0}~~~~,
$$
where $\tilde{0}=(0 \dots 0)^T$. At an equilibrium, the state of the system stays ``at rest" in that phase location, i.e., does not change its present behavior.

To analyze the stability of equilibrium points, we use the Jacobian matrix $D\mathbf{\tilde{f}}(\mathbf{\tilde{x}})$ of the vector field $\mathbf{\tilde{f}}(\mathbf{\tilde{x}})\equiv \frac{\partial\left(f_1, f_2 \dots f_n\right)}{\partial\left(x_1, x_2, \dots , x_n\right)}$ which is a matrix of partial derivatives:
$$
D\tilde{f}(\mathbf{\tilde{x}}) = \begin{bmatrix} 
\frac{\partial f_1}{\partial x_1} & \cdots & \frac{\partial f_1}{\partial x_n} \\
\vdots & \ddots & \vdots \\
\frac{\partial f_n}{\partial x_1} & \cdots & \frac{\partial f_n}{\partial x_n} 
\end{bmatrix}~~~~.
$$
Nature of the eigen values of the Jacobian at the equilibrium point $\mathbf{x}_0$ enables us in determining the stability criteria of the equilibrium. They also speculate the behaviors of trajectories in the neighbourhood of $\mathbf{\tilde{x}}_0$. We sum these properties as :
\begin{itemize}
\item If all eigenvalues of the Jacobian at \( \mathbf{\tilde{x}}_0 \) have non-zero real parts, the equilibrium is hyperbolic.
\begin{itemize}
\item if all the eigen values are consisting negative real parts, we are at a asymptotically stable node. Trajectories are converging to the concerned equilibrium.
\item on the contrary, all real parts are positive indicates towards an unstable node from which the trajectories are repelled.
\item if the set of real parts contains both positive and negative values, they are saddle. Some trajectories are attracted and some are moving away from such points.
\end{itemize}
 \item A point is non-hyperbolic if at least one eigenvalue of the Jacobian matrix at that point has zero real part.
 \begin{itemize}
\item keeping the real parts of some eigen values zero, if the rest eigen values are negative, stable nodes are formed.
\item the upper case turns to be an unstable one when the set of real parts of the eigen values consists of only positive real parts along with at least one vanishing real part.
\item with some zero, both positive and negative real parts arise for saddle points. 
\end{itemize}
\end{itemize}

 Occurrence of non-hyperbolic equilibrium point typically means that linear stability analysis (using the Jacobian) is insufficient for determining the behavior of the system near the equilibrium, and a nonlinear analysis is necessary to understand the system's behavior\cite{book_G.C.Layek}.

For hyperbolic equilibrium points (where all eigenvalues have nonzero real parts), linearization works well. The Hartman-Grobman theorem guarantees that the qualitative behavior of the nonlinear system near the equilibrium is the same as the linearized system. This means that, around a hyperbolic equilibrium, the trajectories of the nonlinear system will behave just like the trajectories of a linear system.

For non-hyperbolic equilibrium points, this is not true. The eigenvalues having zero real parts mean that linearization can only give partial information about the system's dynamics, and more sophisticated techniques must be used to determine the behavior near the equilibrium.

An $f(T)$ gravity model is analyzed in the article \cite{Sadatian_2024} which supports the cyclic universe by dynamical system analysis. Barrow holographic dark energy model within the framework of $f{(Q,~L_m)}$ gravity is studied considering both non-interacting and interacting scenarios in the article \cite{Samaddar_2024}. Kalb-Ramond gravity is studied with the help of a dynamical system to find critical energy levels \cite{Rezeda_2024}. Scalar torsion $f{(T,~\phi)}$ gravity is studied as a dynamical system and critical points are studied \cite{Kadam_2024}. A smooth transition from a deceleration phase to an accelerating expansion is shown in the study of $f(Q,~B)$ gravity's dynamical system analysis, $B$ being the boundary term \cite{Lohakare_2024}. The article \cite{Bhagat_2024} has noticed that within the framework of phase space analysis method $f(Q,~T)$ gravity, the presence of unstable critical points represents the deceleration expansion during the early time (radiation or matter-dominated epochs). Stable critical points indicate accelerated expansion, i.e., deSitter space. Dynamical system approach by defining dimensionless variables and identifying critical points of $f(T, \phi)$ gravity is done in
\cite{kadam2024dynamical}. The paper concludes that scalar field models in $f(T,\phi)$ teleparallel gravity, using both exponential and power-law couplings with an exponential potential, can yield stable critical points that describe various cosmic epochs. These models can support a late-time accelerated universe, consistent with current observations. The dynamical system approach successfully constrains model parameters and confirms the viability of such alternative gravity frameworks for explaining cosmic acceleration. The findings of \cite{duchaniya2023dynamical} suggest that $f(T,\phi)$ gravity models can offer alternative explanations for cosmological phenomena, potentially addressing issues such as the accelerated expansion of the universe. However, the authors note that further research is needed to refine these models and compare their predictions with observational data. The paper \cite{mishra2024scalar} introduces a novel and promising approach to dark energy modeling through the use of fractional calculus in holographic dark energy. This extension of existing models is expected to better describe the universe’s accelerated expansion and could lead to improved cosmological predictions. Further tests and comparisons with observational data are essential to establish the full potential of this model in advancing our understanding of dark energy. The paper \cite{kadam2024teleparallel} aims to explore a modified gravity model that combines teleparallel gravity with quintessence through nonminimal boundary couplings, to better explain the universe’s accelerated expansion. This approach seeks alternatives to the $\Lambda$CDM model by addressing its theoretical issues—like fine-tuning—and offering a dynamic explanation for dark energy using scalar fields that interact more intricately with spacetime geometry. The paper  \cite{kadam2023dynamical} is motivated by the desire to find viable, theoretically motivated models of cosmic acceleration within the richer landscape of modified teleparallel gravity, and to analyze them dynamically to see if they can replicate the observed universe. The article \cite{kadam2022teleparallel} has explored the cosmological dynamics
of dark energy through the prism of scalar-torsion gravity
[36,37] in the context of power-law couplings with the kinetic
term. The study of \cite{pati2022rip} explores the role of dark energy radiation (DER) in cosmology, examining how it affects the universe's expansion rate and comparing various models with current observational data. It also discusses potential particle candidates, such as neutrinos, axions, and dark photons, and their role in DER, as well as their detectability in future experiments. The conclusion suggests that while DER remains undetected with current data, future high-precision experiments could provide crucial insights into its properties. The paper presents a 5-year data release from the Dark Energy Survey Supernova Program, including light curves of 1,635 Type Ia supernovae spanning redshifts from 0.1 to 1.13. It compares two photometric methods, finding that Scene Modelling Photometry (SMP) offers better precision and accuracy than forced PSF photometry\cite{kadam2024constraining}.

In this article, different $f(Q)$ models are chosen. Almost all physical mathematical structures are chosen for dynamical system construction. We expect to portray all the cosmological evolution points via this descriptive analysis.

In the next section, we will present the basic equations related to $f(Q)$ cosmology. After that, six  different $f(Q)$ cosmology models and their respective dynamical systems will be studied. At the end, brief discussion and concluding remarks will be given.
\section{Cosmology with Different Functions $f$ of Nonmetricity Scalar $Q$}
After introducing the Symmetric Teleparallel Gravity, the action of modified gravity, known as $f(Q)$ gravity, is given by\cite{Jim_nez_2018, Jim_nez_2020}
\begin{equation}\label{modified gravity's action}
S=\int{\left[-\frac{1}{16 \pi G }f(Q)+{\cal L}_m\right]\sqrt{-g}d^4 x}~~~~~~~,
\end{equation}
where $g$ is the determinant of the metric tensor $g_{\mu \nu}$, and $ {\cal L}_m$ is Lagrangian density of matter. $f(Q)$ is an arbitrary function of the non-metricity scalar $Q$ which is liable for the gravitational interaction. The non-metricity tensor is given by \cite{Jim_nez_2018}
\begin{center}
$ Q_{\lambda \mu \nu}=\nabla_{\lambda} g_{\mu \nu}-\Gamma^{\beta}_{\lambda \mu}g_{\beta \nu}-\Gamma^{\beta}_{\lambda \nu} g_{\mu \beta}~~~~~~~,$
\end{center}
where $\Gamma_{\mu \nu}^{\beta}$ is the affine connection defined in equation (\ref{Chritoffel_2}) and $\nabla_{\lambda}$ denotes the contravariant differentiation with respect to the coordinate $x^\lambda$.

The non-metricity scalar $Q$ is written as \cite{Khyllep:2022spx}
\begin{equation}
Q= -\frac{1}{4}Q_{\alpha \beta \gamma} Q^{\alpha \beta \gamma}+\frac{1}{2}Q_{\alpha \beta \gamma} Q^{\alpha \beta \gamma}+\frac{1}{4}Q_\alpha Q^\alpha -\frac{1}{2}Q_\alpha \tilde{Q}^\alpha~~~~~~~,
\end{equation}
where $Q_\alpha \equiv Q_{\alpha \mu}^{\mu}$ and $\tilde{Q}^\alpha \equiv Q_\mu^{\\mu \alpha}$ are obtained by considering the non-metricity tensor $Q_{\alpha \mu \nu} \equiv \nabla_\alpha g_{\mu \nu}$ which again defines the non-metricity scalar as $Q=-Q_{\alpha \mu \nu} Q$.

Variating the action \eqref{modified gravity's action} and setting $8 \pi G=1$ for simplicity, the field equations are derived as \cite{Jim_nez_2020, Dialektopoulos_2019}:

$$\frac{2}{\sqrt{-g}}\nabla_\alpha \left\lbrace\sqrt{-g} g_{\beta \nu} f_Q \left[-\frac{1}{2} L^{\alpha \mu \beta} +\frac{1}{4} g^{\mu \beta}\left(Q^\alpha-\tilde{Q}^\alpha \right) -\frac{1}{8} \left(g^{\alpha \mu} Q^{\beta} + g^{\alpha \beta} Q^\mu \right) \right] \right\rbrace~~~~~~~~~~~~~~~~~~~~~~~~~~~~~~~~~~~~~~~~~~$$ 
\begin{equation}
~~~~~~~~~~~~~~~~~~~~~~~~~~~~~~~~~~~~~~~+f_Q \left[-\frac{1}{2} L^{\mu \alpha \beta}-\frac{1}{8}\left(g^{\mu \alpha} Q^\beta + g^{\mu \beta}Q^\alpha \right) + \frac{1}{4} g^{\alpha \beta}\left(Q^\mu - \tilde{Q}^\mu \right) \right] Q_{\nu \alpha \beta} + \frac{1}{2} \delta_\nu^\mu f=T^\mu_\nu~~~~~~~~~~~,
\end{equation}
where $L^\alpha_{\mu \nu}=\frac{1}{2}Q_{\mu \nu}^\alpha - Q^\alpha_{(\mu~\nu)}$ is the disformation tensor, $T_{\mu \nu}$ is the energy-momentum tensor and $f_Q \equiv \frac{d f }{d Q}$.

At the background level, consider a homogeneous, isotropic and spatially flat Friedmann-Lemaitre-Robertson-Walker (FLRW) space time whose metric form is given by
\begin{equation}
ds^2 = -dt^2+a^2(t)(dx^2+dy^2+dz^2)~~~~~~~~~,
\end{equation}
where $t$ is the cosmic time, $a(t)$ is the scale factor and $x,~y,~z$ are the cartesian coordinates. For the non-metricity scalar, we obtain $ Q=6H^2$ in FLRW spacetime metric, where $H=\frac{\dot{a}}{a}$ is the Hubble parameter and the upper dot denotes the derivative w.r.t. $t$. Imposing the splitting $f(Q)= Q + F(Q)$ and by using the FLRW metric, the corresponding Friedmann equations turn,
\begin{equation}\label{Friedmann density equation}
3H^2=\rho + \frac{F}{2} -Q F_Q~~~~~~~~~~~~~~and
\end{equation}
\begin{equation}\label{Friedmann pressure equation}
(2QF_{QQ}+F_Q+1)\dot{H}+\frac{1}{4}(Q+2QF_Q-F)=-2p~~~~~~~~~~~~,
\end{equation}
with $F_Q \equiv \frac{dF}{dQ},~~F_{QQ} \equiv \frac{d^2 F}{dQ^2}$, where $\rho$ and $p$ respectively are the energy density and pressure of the fluid present in the model. Then the conservation equation turns
\begin{equation}\label{Conservation equation}
\dot{\rho}+3H(1+\omega)\rho=0~~~~~~~~,
\end{equation}
with $\omega \equiv \frac{p}{\rho}$, the equation of state for the fluid content.

The effective total energy density $\rho_{eff}$, pressure $p_{eff}$ and the corresponding equation of state $\omega_{eff}$ respectively can be presented as 
\begin{equation}
\rho_{eff} \equiv \rho+\frac{F}{2}-QF_Q,
\end{equation}
\begin{equation}
p_{eff} \equiv \frac{\rho(1+\omega)}{2QF_{QQ}+F_Q+1}-\frac{Q}{2}~~~~~~and
\end{equation}
\begin{equation}
\omega \equiv \frac{p_{eff}}{\rho_{eff}}=-1+\frac{\Omega_m(1+\omega)}{2QF_{QQ}
+F_Q+1}~~~~.
\end{equation}
For an accelerating universe, $\omega_{eff}<-\frac{1}{3}$ and the energy density parameters $\Omega_m$ and $\Omega_Q$ are given by
\begin{equation}
\Omega_m=\frac{\rho}{3H^2}~~~~~~and ~~~~~\Omega_Q=\frac{\frac{F}{2}-QF_Q}{3H^2}~~~~~~~.
\end{equation}
The first Friedmann equation \eqref{Friedmann density equation} becomes simply 
\begin{equation}
\Omega_m + \Omega_Q=1~~~~~~~.
\end{equation}

We consider the linear perturbation level and concentrate on the matter density contrast $\delta=\frac{\delta_\rho}{\rho}$, where $\delta_\rho$ is the perturbation of the matter energy density. Particularly, the evolution equation of the matter over density at the quasi-static limit evolves as \cite{Jim_nez_2020, Anagnostopoulos_2021}
\begin{equation}\label{Quasi static equation}
\ddot{\delta}+2 H \dot{\delta}=\frac{\rho \delta}{2(1+F_Q)}~~~~~~~,
\end{equation}
where the denominator in the RHS explains the appearance of Newton’s constant. The temporal derivative terms in the perturbed equations are neglected in the quasi-static limit at tiny scales. Just spatial derivative terms are sustained \cite{shah2019, Anagnostopoulos_2023, Lazkoz:2019sjl}.

In the next section, we will construct the autonomous system for our models and describe the related phase portraits.

  \section{Dynamical System Analysis}
In this section, for a general function $F(Q)$, the dynamical system of the background and perturbed equations will be constructed. In this aspect, equations \eqref{Friedmann density equation}, \eqref{Friedmann pressure equation}, \eqref{Conservation equation} and \eqref{Quasi static equation} transform into first order autonomous system by taking the dynamical variables into consideration:
\begin{equation}\label{Cosmological equations}
x=\frac{F}{6H^2},~~~y=-2F_Q,~~~u=\frac{d(ln~ \delta)}{d(ln~ a)}~~~~~~.
\end{equation}

As a result, while variables $x$ and $y$ are associated to universe's background evolution, the variable $u$, on the contrary, estimates the expansion of matter perturbations. Hence, the matter density contrast is positive at any time. $ u>0$ indicates the expansion of matter perturbations and $u < 0$ denotes their decay, provided the matter density contrast $\delta$ stays positive.

The cosmic background parameters $ \Omega_m$, $\Omega_Q$ and $\omega_{eff} $ are given by\cite{Jim_nez_2020} :
\begin{equation}
\Omega_m=1-x-y~~~~~~~~,
\end{equation}
\begin{equation}
\Omega_Q=x+y~~~~~~~~~~ and
\end{equation}
\begin{equation}
\omega_{eff}=-1+\frac{(1-x-y)(1+\omega)}{2QF_{QQ}-\frac{y}{2}+1}~~~~~~~.
\end{equation}

The cosmological equations can now be expressed as the following autonomous system in terms of variables \eqref{Cosmological equations} as:

\begin{equation}\label{x_derivative}
x^{\prime}= -\frac{\dot{H}}{H^2}(y+2x)~~~~~~~~~~,
\end{equation}
\begin{equation}\label{y_derivative}
y^{\prime}=2y \frac{\dot{H}}{H^2} \frac{QF_{QQ}}{F_Q}~~~~~~~~~~~and
\end{equation}
\begin{equation}\label{u_derivative}
u^{\prime}=-u(u+2)- \frac{3(x+y-1)}{2-y}-\frac{\dot{H}}{H^2}u~~~~~~~~~,
\end{equation}
where $(^\prime)$ denotes the differentiation with respect to $ln(a)$ and $(\dot{})$ denotes the differentiation with respect to $t$ and
\begin{equation}
\frac{\dot{H}}{H^2}=\frac{3(x+y-1)}{4QF_{QQ}-y+2}~~~~~~~.
\end{equation}
The physical system is a product space of the perturbed space $\mathbb{P}$ which is a function of  the variable $u$, and the background phase space $\mathbb{B}$, which is a function of the variable $x$ and $y$. The phase space under the physical condition $0 \leq \Omega_m \leq 1$ of the combined system is 
\begin{equation}
 \Psi=\mathbb{B} \times \mathbb{P}=\left\{(x,y,u) \in\mathbb{R}^2 \times \mathbb{R} : 0 \leq x+y \leq 1\right\}~~~~~~.
\end{equation}
Note that the projection of orbits of the product space $\Psi$ on space $\mathbb{B}$ reduces to the corresponding background orbits.
\subsection{\large {\bf Model I :} $F(Q) = exp\left\{nQ\right\}$}
Here $n$ is a parameter. This model supports the Big Bang Nucleosynthesis (BBN) requirement\cite{Anagnostopoulos_2021} and able to explain the late time cosmic acceleration. This model can mimic dark energy behavior, supports acceleration of the expansion of the universe. It is a simple yet non-trivial extension of GR (which corresponds to $f(Q)=Q$) \cite{Narawade2025}.

The exponential form allows for analytic tractability in cosmological equations. $F_Q(Q)$ is derived as :
\begin{equation}
F_Q (Q)=n e^{nQ} ~~~~~.
\end{equation}

Then with the help of \eqref{Cosmological equations},  $Q F_{QQ}(Q)$ is given by 
\begin{equation}
Q F_{QQ}(Q)=n^2 Q e^{nQ}=-\frac{y}{2}\ln\left(- \frac{y}{2 n}\right)~~~~.
\end{equation}

From the equations \eqref{x_derivative}, \eqref{y_derivative} and \eqref{u_derivative}, we obtain

\begin{equation}
x^{\prime}=\frac{-3(x+y-1)(2x+y)}{2yln(- \frac{y}{2 n})+y-2}~~~~,
\end{equation}

\begin{equation}
y^{\prime} = \frac{24 y (x+y-1)ln(- \frac{y}{2 n})}{2yln(- \frac{y}{2 n})+y-2}~~~~and
\end{equation}

\begin{equation}
u^{\prime} = -u(u+2)- \frac{3(x+y-1)}{2-y}-\frac{3(x+y-1)u}{2yln(- \frac{y}{2 n})+y-2}~~~~.
\end{equation}

Four critical points are obtained by solving 
$\begin{pmatrix}
{x}^\prime \\
{y}^\prime \\
{u}^\prime
\end{pmatrix}
= \tilde{0}$. These critical points, the eigen values of the Jacobian matrix at those points and some other corresponding values are listed in the table-1 :

\begin{table}[h!]\label{Model 1}
\caption{Critical points, eigen values of the Jacobian at those points, EoS parameter, expansion of matter disturbances and  stability conditions for $F(Q) = exp\left\{nQ\right\}$ model.}

\centering
 \begin{tabular}{||c c c c c c c c||} 
 \hline
 Critical points & Eigen-values & $ \Omega_m$ & $ \Omega_Q$ & $ \omega_{eff}$ & $u$ & Equilibrium type & Stability condition. \\ [0.5ex] 
 \hline
 $\left(n,~-2n,~\frac{-7-\sqrt{73}}{4}\right)$ & $\left(-6,~-24, ~ \frac{\sqrt{73}}{{2}}\right)$ & $1+n$ & $-n$ & $\omega$ & $ \frac{\sqrt{73}}{{2}}$& hyperbolic & Unstable \\ 
 $\left(n,~ -2n,~ \frac{-7+\sqrt{73}}{4}\right)$  & $\left(-6,~ -24,~-\frac{\sqrt{73}}{{2}}\right)$ & $1+n$ & $-n$ & $\omega$ & $ -\frac{\sqrt{73}}{{2}}$& hyperbolic & Stable  \\
 $(1+2n,~ -2n,~ -2)$ & $(-1,~ 0,~ 2)$ & $0$ & $1$ & $-1$ & $-2$& non-hyperbolic & Saddle \\
 $(1+2n,~ -2n,~ 0)$ & $(-2,~ -1,~ 0)$ & $0$ & $1$ & $-1$ & $-2$& non-hyperbolic & Stable \\[1ex] 
 \hline
 \end{tabular}
\end{table}

\begin{figure}[h!]
$~~~~~~~~~~~~~~~~~~Fig.1(a)~~~~~~~~~~~~~~~~~~~~~~~~~~~~~~~~~~~~~~Fig.1(b)~~~~~~~~~~~~~~~~~~~~~~~~~~~~~~~~~~~~~~~~Fig.1(c)~~~$\\
\includegraphics[scale=0.41]{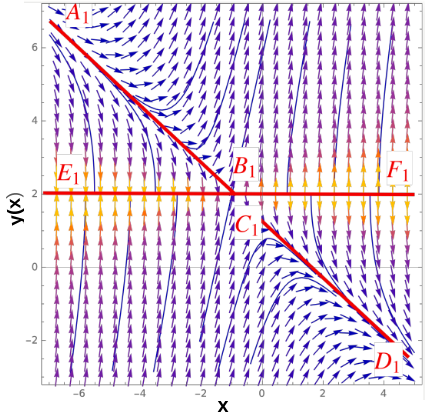}
\includegraphics[scale=0.38]{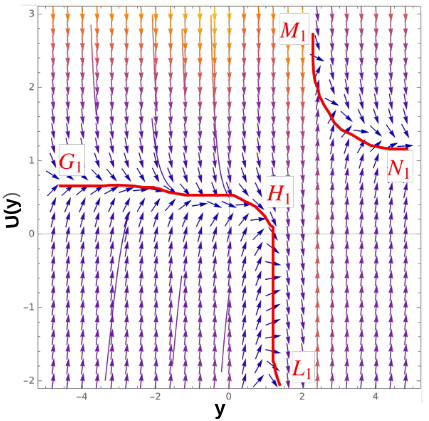}
\includegraphics[scale=0.38]{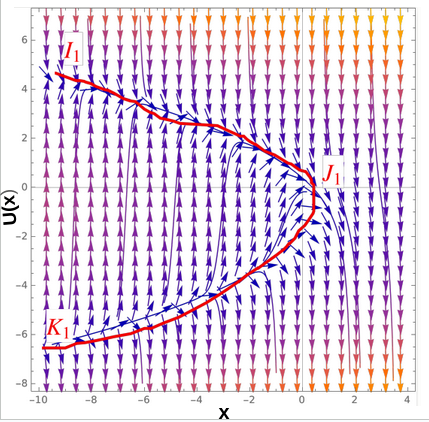}\\
\centering{Fig. 1(a)-1(c) : Phase portraits for the cosmological system obeying the non-metricity function $F(Q) = exp\left\{nQ\right\}$. Fig. $1(a),~1(b)~and~1(c)$ represent $y$ vs $x$, $u$ vs $y$ and $u$ vs $x$ spaces respectively.}
\end{figure}

{\bf \large Critical Point $\left(n,~-2n,~-\frac{7+\sqrt{73}}{4}\right)$ :} This point's curve is a solution which shows negative contribution of dark energy component. Late time cosmic acceleration is having the same equation of state as the fluid's effective equation of state. Through this curve, effective equation of state behaves like the same of a dynamical dark energy model, specifically not of a cosmological constant type except at $n=1$. $u=-\frac{\sqrt{73}+7}{4}$ shows decay in perturbation. The one-dimensional curve possesses one positive and two negative eigen values of the Jacobian matrix which indicates that the critical point is unstable in nature. This point is unable to explain the formulation of structures at the perturbation levels. Hence the matter over density $\delta$ fluctuates as $ \rho \propto a(t)^{-\frac{7+\sqrt{73}}{4}}$.

The figure $(1a)$ shows $y(x)$ vs $x$ phase plane in which two curves $A_1B_1$ and $B_1F_1$ repel the arrows towards the respective directions of $\overrightarrow{A_1B_1}$ and almost in a perpendicular direction to $B_1F_1$ respectively them. So the curves act like sources. Similarly, by considering the phase plane $u(x)$ vs $x$ in figure $(1c)$, the curve $K_1J_1$ also repels the arrows towards its direction and acts like a source. However, in the phase plane $u(y)$ vs $y$ in figure $(1b)$, there exists a curve $GH$, which attracts the arrows and acts like a sink.

The existence of a dynamical repulsive point in cosmology implies the formation and evolution of cosmic structures like galaxies and clusters. The repulsive force it exerts could potentially influence the gravitational attraction between matter, affecting the way these structures form and develop.

The implications of a dynamical repulsive point is also its potential to explain the observed cosmic acceleration. This phenomenon indicates that the expansion of the universe is not slowing down rather is expanding. While the prevailing explanation for this acceleration is dark energy a dynamical repulsive point could offer an alternative or complementary explanation towards the observations indicating late time cosmic acceleration.\\

{\bf \large Critical Point $\left(n, -2n, \frac{-7+\sqrt{73}}{4}\right)$ :} This critical point is also a curve which shows effective equation of state equal to the same of dark energy chosen. Matter perturbation increases as $u=\frac{-7+\sqrt{73}}{4} > 0$. This critical point has three negative eigen values ensuring the point to be stable in nature. It also shows the negative contribution of dark energy components. Its behavior at late time cosmic acceleration is again same with the effective EoS.

$y$ vs $x$ phase plane in figure $(1a)$ shows attracting points along the curves $C_1D_1$ and $E_1B_1$ which attracts towards the directions of $C_1D_1$ and perpendicularly towards $E_1B_1$ respectively. 

In $u$ vs $y$ phase plane in figure $(1b)$, there exists a attractor curve $G_1H_1$ which attracts the arrows along the direction of $G_1H_1$, as mentioned before.

In $u$ vs $x$ phase plane in figure $(1c)$, $I_1J_1$ is the attractor curve that acts like a sink.\\

{\bf \large Critical Point $(1+2n,~-2n,~-2)$ :} These fixed points are non-hyperbolic. This point's curve is a solution which shows positive contribution of effective dark energy. The effective Equation of state is same with the Equation of state of the cosmological constant dominated universe. 

The eigen values of the Jacobian matrix at this point are $-1,~0~$ and $2$, so the point always be a saddle one.

Since $u=-2$, thus the matter density $\delta$ varies as $a^{-2}$. 

First, we should follow that the eigen values are free of $n$. Hence the hyperbolicity at this critical point does not depend on any free parameter. In $u(x)$ vs $x$ phase plane in figure $(1c)$, $J_1$ is the only one saddle point. For $u(y)$ vs $y$ phase plane in $(1b)$, $H_1L_1$ acts as a saddle line. Also for $y(x)$ vs $x$ phase plane, in figure $(1a)$, $B_1$ is the saddle point. Some fixed points may be non-hyperbolic, especially when the potential has a plateau or when the universe asymptotically approaches deSitter space ($\omega_{eff}\rightarrow-1$).\\

{\bf \large Critical Point $(1+2n,~-2n,~0)$ :} This is another set of non-hyperbolic critical point. Each points on this curve also leads to a solution dominated by positive dark energy components as $\Omega_Q=1$. The effective equation of state also behaves like the same of the cosmological constant. At the perturbation level, $u=0$, which implies that the matter perturbation remains constant. Cosmologically, this non-hyperbolic type fixed point resembles the last one, i.e., this is pointing towards a $\Lambda$CDM model where we feel lucky knowing the cosmic coincidence of fractional density contributions of different constituents of cosmos.\cite{Overduin1997}

The eigen values of the Jacobian matrix at this critical point are $-2,~-1~\text{and}~0$. So the curve is non-hyperbolic. Though these eigen values are parameter independent.

Phase portrait analysis for this critical point resembles that of the second one.

In this case, $x$ is the central variable and $y,~u$ are the stable variables. The
corresponding matrices are $\tilde{A}= \tilde{0}$ and $\tilde{B}= \begin{pmatrix}
-1 & 0\\  0 & -2
\end{pmatrix}$. The center manifold has now the form $y=h_1(x)$, and $u=h_2(x)$; the approximation $N$ has two components :

$$ N_1(h_1(x))=h_1^{\prime}(x)x^{\prime}-y^{\prime} = h_1^{\prime}(x) \frac{-3(x+y-1)(2x+y)}{2yln(- \frac{y}{2 n})+y-2} - \frac{24 y (x+y-1)ln(- \frac{y}{2 n})}{2yln(- \frac{y}{2 n})+y-2}$$\\

and $$ N_2(h_2(x))=h_2^{\prime}(x)x^{\prime}-u^{\prime} = h_2^{\prime}(x) \frac{-3(x+y-1)(2x+y)}{2yln(- \frac{y}{2 n})+y-2} + u(u+2) + \frac{3(x+y-1)}{y-2}+\frac{3(x+y-1)}{2yln(- \frac{y}{2 n})+y-2}u~~~~~.$$

For Zeroth approximation,

$$ N_1(h_1(x)) = 0 +o(x^2)~~\text{and}~~ N_2(h_2(x)) = \frac{3}{2}(x-1)+o(x^2)~~~.
$$
The central manifold theory is used in the context of dynamical systems to simplify the analysis of complex systems by focusing on the behavior near an equilibrium point. For exponential models which often exhibit nonlinear and possibly chaotic dynamics, it provides a way to reduce the system to a lower-dimensional space that captures the core behavior close to the equilibrium.

\subsection{\large {\bf Model II :} $F(Q) = Q+\eta log_e(\alpha Q)$ }
Here $\eta$ and $\alpha$ are any two arbitrary parameters as defined in \cite{Sanjay_2020}. There is a theoretical basis particularly embedded in this model. Addition of $log(\alpha Q)$  with the standard $f(Q)=Q$ , i.e., the GR equivalent trivial term introduces nontrivial dynamics which is able to mimic the cosmological constant at late times. This occurs due to the small value of Hubble's parameter $Q\sim H^2$. As the logarithmic term grows slowly without any noticable contribution to the equation of motion , accelerated expansion can be mimiced without any dark energy model's introduction \cite{Narawade2024}.

For Model II,
\begin{equation}
F_Q (Q)=1+\frac{\eta}{Q}=-\frac{y}{2} ~~~~\text{and}
\end{equation}

\begin{equation}
Q F_{QQ}(Q)=-\frac{\eta}{Q}=1+\frac{y}{2}~.
\end{equation}
Again from the equations \eqref{x_derivative}, \eqref{y_derivative} and \eqref{u_derivative} we get,
\begin{equation}
x^{\prime}=\frac{-3(x+y-1)(2x+y)}{y+6}~~~,
\end{equation}

\begin{equation}
y^{\prime} = \frac{-6 (x+y-1)(2+y)}{y+6}~~~~and
\end{equation}

\begin{equation}
u^{\prime} = -u(u+2)- \frac{3(x+y-1)}{2-y}-\frac{3(x+y-1)u}{y+6}~~.
\end{equation}

Vanishing the derivative $w.r.t~(log_e a)$ of the vector 
$\begin{pmatrix}
x \\ y \\ u
\end{pmatrix}$
leads us to find four critical points which are tabulated in table 2.

\begin{table}[h!]
\caption{Critical points, eigen values of the Jacobian, EoS parameter, expansion of matter disturbances and  stability conditions for $F(Q) = Q+\eta log_e(\alpha Q)$.}
\centering
 \begin{tabular}{||c c c c c c c c||} 
 \hline
  Critical points & Eigen-values & $ \Omega_m$ & $ \Omega_Q$ & $ \omega_{eff}$ & $u$ & Equilibrium type & Stability condition. \\ [0.5ex] 
 \hline
 $(1-y,~y,~-2)$ & $(0,~2,~3)$ & $0$ & $1$ & $-1$ & $-2$ & non-hyperbolic & Unstable \\ 
 $(1-y,~y,~0)$  & $(-2,~0,~3)$ & $0$ & $1$ & $-1$ & $0$ & non-hyperbolic & Saddle  \\
 $(1,~-2,~-\frac{3}{2})$ & $(-3,~-3,~\frac{5}{2})$ & $2$ & $-1$ & $\omega$ & $-\frac{3}{2}$& hyperbolic & Unstable \\
 $(1,~-2,~1)$ & $(-3,~-3,~-\frac{5}{2})$ & $2$ & $-1$ & $\omega$ & $1$ &hyperbolic & Stable \\[1ex] 
 \hline
 \end{tabular}
\end{table}

\begin{figure}[h!]
\begin{center}
$~~~~~~~~~~Fig.~2(a)~~~~~~~~~~~~~~~~~~~~~~~~~~~~~~~~Fig.~2(b)~~~~~~~~~~~~~~~~~~~~~~~~~~~~~~Fig.~2(c)~~~$\\

\includegraphics[scale=0.33]{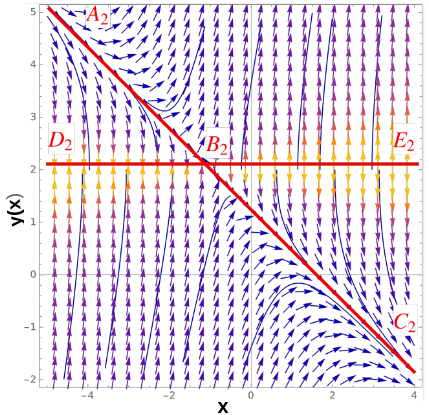}
\includegraphics[scale=0.32]{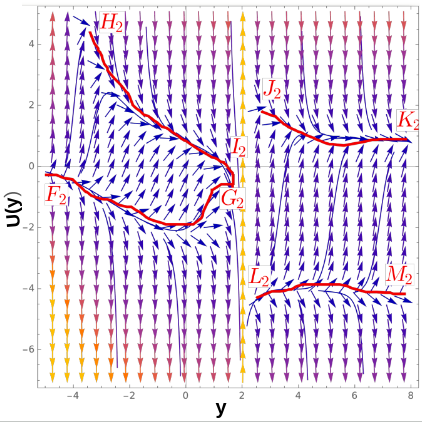}
\includegraphics[scale=0.35]{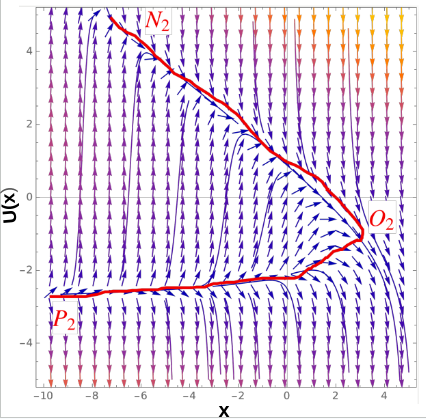}\\
\centering{Fig.2(a)-2(c) : Phase portraits for the cosmological system obeying the non-metricity function $F(Q) = Q+\eta log_e(\alpha Q)$. Fig. $2(a)$, $2(b)$ and $2(c)$ represent $y$ vs $x$, $u$ vs $y$ and $u$ vs $x$ spaces respectively.}
\end{center}
\end{figure}

{\bf \large Critical Point $(1-y,~y,~-2)$ :} This non-hyperbolic critical point denotes an unstable critical point where universe is cosmological constant dominated. Matter evolution is followed to decay as $u_c=-2<0$. Dark energy component is followed to occupy the whole fraction of constituent energy density. In the phase portraits figure $2a-2c$, we find unstable lines like $A_2B_2$, $B_2 E_2$, $F_2 G_2$, $L_2 M_2$ and $P_2 O_2$. 

As the universe evolves, certain states (like deSitter space) become late-time attractors, where the system tends toward them as time progresses, regardless of initial conditions. Non-hyperbolic equilibrium points are often related to these late-time solutions. For example, in the case of a universe dominated by dark energy (e.g., a cosmological constant or a slowly rolling scalar field), the system's dynamics might approach a deSitter space where the universe expands exponentially at late times. The equilibrium point corresponding to deSitter space might be non-hyperbolic, meaning that it's a delicate balance between different energy components, and its stability could depend on the detailed properties of the scalar field and potential. Small variations can influence the rate of expansion, possibly leading to transient behaviors before the universe settles into a final equilibrium.

{\bf \large Critical Point $(1-y,~y,~0)$ :} Second point in this row is a non-hyperbolic saddle with cosmological constant like effective equation of state. Here also matter contribution is null. Matter evolution is found to be stable as $u$ vanishes at this critical point. The points $B_2$ and $O_2$ are indications of such saddles.

At late times in the universe’s history, the cosmological constant $(\Lambda)$ or dark energy dominates the expansion. The transition to this regime from matter or radiation domination can involve non-hyperbolic points in the dynamical system of the cosmological equations. The universe may evolve toward a deSitter solution, and the potential non-hyperbolic nature of the equilibrium near w=-1 (where w is the equation of state of dark energy) makes the stability of such a point less straightforward. \cite{Harrison2000}

{\bf \large Critical Point $(1,~-2,~-\frac{3}{2})$ :} Third critical point represents an unstable equilibrium point where the effective equation of state resembles that of the dynamical dark energy component present in the model. Matter evolution is decaying.

{\bf \large Critical Point $(1,~-2,~1)$ :} The ultimate one is a stable critical point. Matter is increasing as $u=1>0$. This is following the equation of state of the dynamical dark energy constituent of our model. $B_2 C_2$, $H_2 I_2$, $J_2 K_2$ and $N_2 O_2$. \\ 


\subsection{\large {\bf  Model III :} $F(Q) = Q+\eta Q^{-1}$ }

Here $\eta$ is an arbitrary parameter\cite{Jim_nez_2020}.  The extra nonlinear term $\frac{\eta}{Q}$ contributes a cosmological fluid, EoS of which mimics the effects of dark energy without the need of cosmological constant. This is a dynamical alternative to dark energy driven by the modification of gravity itself. This model also works towards the singularity avoidance \cite{Shabani2024}. Equations of this model are constructed as follows
\begin{equation}
F_Q (Q)=1+\frac{\eta}{Q^2}=-\frac{y}{2} ~~~~\text{and}
\end{equation}

\begin{equation}
Q F_{QQ}(Q)=-\frac{2 \eta}{Q^2}=y+2
\end{equation}

Again from the equations \eqref{x_derivative}, \eqref{y_derivative} and \eqref{u_derivative}, we get,

\begin{equation}
x^{\prime}=\frac{-3(x+y-1)(2x+y)}{3y+10}~~,
\end{equation}

\begin{equation}
y^{\prime} = \frac{-12 (x+y-1)(2+y)}{3y+10}~~~~and
\end{equation}

\begin{equation}
u^{\prime} = -u(u+2)- \frac{3(x+y-1)}{2-y}-\frac{3(x+y-1)u}{3y+10}~~~~~~.
\end{equation}

$\begin{pmatrix}
{x}^\prime \\
{y}^\prime \\
{u}^\prime
\end{pmatrix}
= \tilde{0}$, where $\tilde{0}$ is the null vector. Solving this, we can get the critical points are tabulated in table-3.

\begin{table}[h!]
\caption{Critical points, eigen values of the Jacobian, equation of state parameter, expansion of matter disturbances and  stability conditions for $F(Q) = Q+\eta Q^{-1}$.} 

\centering
\begin{tabular}{||c c c c c c c c||} 
 \hline
  Critical points & Eigen-values & $ \Omega_m$ & $ \Omega_Q$ & $ \omega_{eff}$ & $u$ & Equilibrium type & Stability condition. \\ [0.5ex] \hline
 $(1-y,~y,~-2)$ & $(0,~2,~3)$ & $0$ & $1$ & $-1$ & $-2$ & non-hyperbolic & Unstable \\ 
 $(1-y,~y,~0)$  & $(-2,~0,~3)$ & $0$ & $1$ & $-1$ & $0$ &non-hyperbolic & Saddle  \\
 $(1,~-2,~-\frac{3}{2})$ & $(-3,~\frac{5}{2}, 6)$ & $2$ & $-1$ & $\omega$ & $\frac{-3}{2}$& hyperbolic & Unstable \\
 $(1,~-2,~1)$ & $(-3,~\frac{-5}{2}, 6)$ & $2$ & $-1$ & $\omega$ & 1 & hyperbolic & Unstable \\[1ex] 
 \hline
 \end{tabular}
\end{table}

\begin{figure}[h!]
$~~~~~~~~~~~~~~~~~~~~~~~~~Fig-3(a)~~~~~~~~~~~~~~~~~~~~~~~~~~~~~~~~~~~~~Fig-3(b)~~~~~~~~~~~~~~~~~~~~~~~~~~~~~~~Fig-3(c)~~~$

\begin{center}
\includegraphics[scale=0.36]{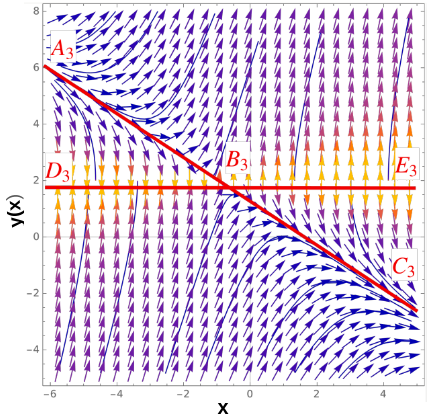}
\includegraphics[scale=0.36]{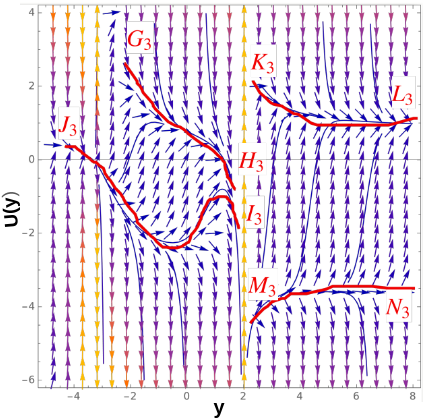}
\includegraphics[scale=0.38]{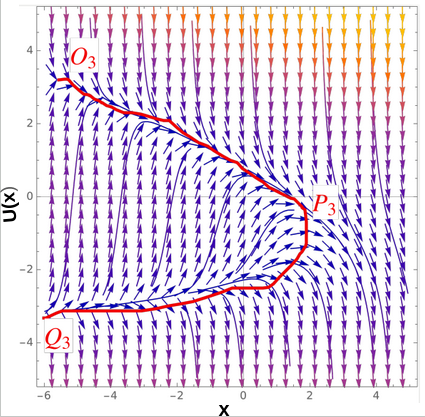}\\
\centering{Fig. 3(a)-3(c) : Phase portraits for the cosmological system obeying the non-metricity function  $F(Q) = Q+\eta Q^{-1}$. $3(a)$, $3(b)$ and $3(c)$ represent $y$ vs $x$, $u$ vs $y$ and $u$ vs $x$ spaces respectively.}

\end{center}
\end{figure}

{\bf \large Critical Point $(1-y,~y,~-2)$ :} This non-hyperbolic type critical point denotes an unstable critical point. Its equation of state is also the same as the $\Lambda CDM$ model. Matter evolution is also followed to decay as $u=-2<0$. Matter contribution is null while the dark energy component occupies the whole fraction of energy density. In this phase portraits $3(a)-3(c)$, we get the unstable curves like $B_3C_3$, $B_3E_3$, $J_3I_3$, $M_3N_3$ and $Q_3P_3$. Stable critical points of this cosmological dynamical system are essential for understanding the universe's long-term behavior, stability, and ultimate fate.

If the cosmological constant ($\Lambda$) dominates the universe’s energy density, the expansion accelerates and eventually reaches a regime where the Hubble parameter $H(t)=\frac{\dot{a}}{a}$ approaches a constant value. At this stage, the universe reaches a non-hyperbolic fixed point, where the expansion rate remains constant. This is because the energy associated with $\Lambda$ eads to a constant pressure and energy density, implying no further acceleration or deceleration. To understand this better, consider a phase space where we track variables such as the scale factor  $a(t)$ and its time derivative $\dot{a}(t)$. The dynamics of the universe can be described by a set of first-order differential equations. In such a system, a fixed point could correspond to a solution where the universe has a constant rate of expansion $\dot{a}=a(t)H(t)$. Another possibility arises when the matter density and the cosmological constant are in a balance that leads to a stable or neutral evolution.

In cosmology, the point where the influence of $\Lambda$ dominates over matter and radiation is often referred to as the ``deSitter point". The deSitter space is a solution to Einstein’s equations with $\Lambda>0$ and no matter. It represents a universe that expands exponentially at late times, driven by the cosmological constant.

{\bf \large Critical Point $(1-y,~y,~0)$ :} The second critical point of this model is a non-hyperbolic saddle point. Its equation of state matches cosmological constant. Non-hyperbolicity arises out of this fact. Here also matter contribution is zero and dark energy takes the whole part of constituent energy density. $B_3$ and $O_3$ are the saddle points. Saddle critical points in inverse cosmological dynamical systems are essential for understanding the universe's evolution, stability, and potential for chaotic behavior.

{\bf \large Critical Point $(1,~-2,~-\frac{3}{2})$ :}  The third critical point is also an unstable critical point. But it's EoS is the same as the dynamical dark energy component present in the model. The phase portrait for this critical point resembles that of the first one. The matter over density $\delta$ fluctuates as $\rho \propto a^{-\frac{3}{2}}$.

{\bf \large Critical Point $(1,~-2,~1)$ :} The final one is also an unstable critical point with the same interpretation as the third one except for the matter density fluctuations.\\


\subsection{\large {\bf Model IV :} $F(Q) = Q+\eta Q^{2}$ }
Here $\eta$ is an arbitrary parameter \cite{Jim_nez_2020}. The extra term of this model represents quantum loop corrections, higher-order curvature terms, nonlinear kinetic or potential terms etc \cite{Harrison2000}.
\begin{equation}
F_Q (Q)=1+2 \eta Q=-\frac{y}{2}~~~\text{and} 
\end{equation}

\begin{equation}
Q F_{QQ}(Q)=2 \eta Q=-\frac{y}{2} -1~~~~.
\end{equation}

Again from the equations \eqref{x_derivative}, \eqref{y_derivative} and \eqref{u_derivative}, we get,

\begin{equation}
x^{\prime}=\frac{(x+y-1)(2x+y)}{y}~~~~,
\end{equation}

\begin{equation}
y^{\prime} = \frac{-2 (x+y-1)(2+y)}{y}~~~~~~~~and
\end{equation}

\begin{equation}
u^{\prime} = -u(u+2)- \frac{3(x+y-1)}{2-y}-\frac{(x+y-1)u}{y}~~~~~~.
\end{equation}
 Similarly, solving $\begin{pmatrix}
{x}^\prime \\
{y}^\prime \\
{u}^\prime
\end{pmatrix}
= \tilde{0}$, where $\tilde{0}$ is the null vector, we can get the critical points, eigen values which are tabulated in table-4.
 
\begin{table}[h!] 
\caption{Critical points, eigen values of the Jacobian, EoS parameter, expansion of matter disturbances and  stability conditions for $F(Q) = Q+\eta Q^{2}$.}
\centering
 \begin{tabular}{||c c c c c c c c||} 
 \hline
  Critical points & Eigen-values & $ \Omega_m$ & $ \Omega_Q$ & $ \omega_{eff}$ & $u$ & Equilibrium type & Stability condition. \\ [0.5ex] \hline
 $(1-y,~y,~-2)$ & $(0,~2,~-\frac{6+y}{y})$ & $0$ & $1$ & $-1$ & $-2$& non-hyperbolic & Unstable for $y\geq-6$ and \\
  &  &  &  &  &  & &Saddle for $y<-6$ \\ 
 $(1-y,~y,~0)$  & $(0,~-2,~-\frac{6+y}{y})$ & $0$ & $1$ & $-1$ & $0$& non-hyperbolic & Stable for $y\leq -6$ and \\
  &  &  &  &  &  & &Saddle for $y>-6$ \\
 $(1,~-2,~\frac{1}{2}(-1-\sqrt{7}))$ & $(\sqrt{7},~-2,~-2)$ & $2$ & $-1$ & $\omega$ & $\frac{1}{2}(-1-\sqrt{7})$& hyperbolic & Unstable \\
 $(1,~-2,~\frac{1}{2}(-1+\sqrt{7}))$ & $(-\sqrt{7},~-2,~-2)$ & $2$ & $-1$ & $\omega$ & $\frac{1}{2}(-1+\sqrt{7})$& hyperbolic & Stable \\[1ex] 
 \hline
 \end{tabular}

\end{table}

\begin{figure}[h!]
$~~~~~~~~~~~~~~~~~~~~Fig-4(a)~~~~~~~~~~~~~~~~~~~~~~~~~~~~~~~~~~~~~~~~Fig-4(b)~~~~~~~~~~~~~~~~~~~~~~~~~~~~~~~~~~~~~~~Fig-4(c)~~~$
\begin{center}
\includegraphics[scale=0.39]{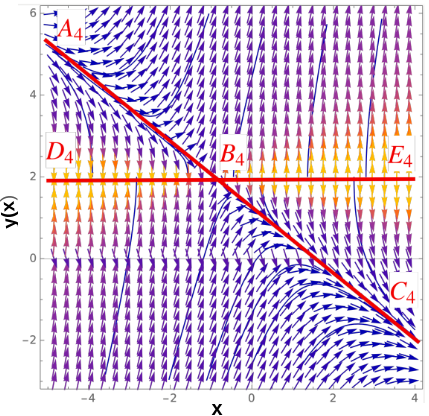}
\includegraphics[scale=0.39]{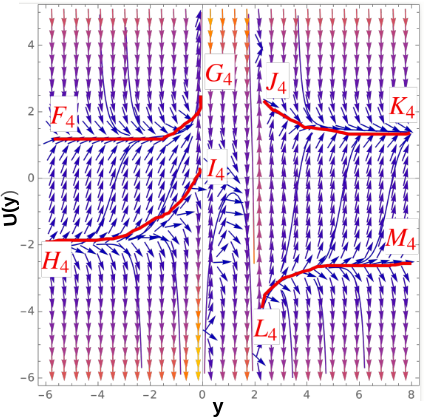}
\includegraphics[scale=0.37]{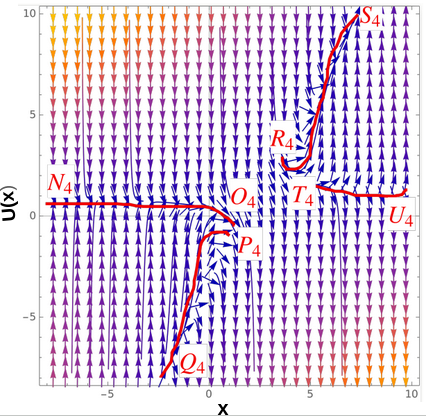}
\centering{Fig.$4(a)$-$4(c)$ Phase portraits for the cosmological system obeying the non-metricity function $F(Q) = Q+\eta Q^{2}$. Figure $4(a)$, $4(b)$ and $4(c)$ represent $y$ vs $x$, $u$ vs $y$ and $u$ vs $x$ spaces respectively.}
\end{center}
\end{figure}

{\bf \large Critical Point $(1-y,~y,~-2)$ :} The first critical point possesses the eigenvalues which depend on $y$. This is essentially a non-hyperbolic type of fixed point. It contains two types of stability conditions. One is unstable for $y\geq -6$ and the other is saddle $y<-6$. In the phase portraits fig. $4(a)$-$4(c)$, $A_4B_4$, $B_4E_4$, $H_4I_4$, $P_4Q_4$ and $T_4U_4$ are unstable curves and $B_4$ is the saddle point. If the universe is near a saddle point, it may exhibit specific patterns in the distribution of matter and energy that can be detected through observations. Its equation of state is the same as the cosmological constant.

{\bf \large Critical Point $(1-y,~y,~0)$ :} Similarly, for the second critical point, it possesses $y$ dependent eigen values. Non-hyperbolicity of this model is independent of any parameter. The model is stable for $y \leq-6$ and saddle for $y>-6$. Stable curves in the phase portraits are $B_4C_4$, $D_4E_4$, $F_4G_4$, $J_4K_4$, $R_4S_4$ and $N_4O_4$. The equation of state is also the same as the cosmological constant.

The cosmological constant $\Lambda$ as a non-hyperbolic fixed point refers to a phase in the universe's evolution where the expansion becomes stable and asymptotically constant. The fixed point itself represents a state of neutral stability, where the universe’s expansion rate does not change, driven by the energy associated with $\Lambda$. This concept is central to understanding the current and future accelerated expansion of the universe under the influence of dark energy.

{\bf \large Critical Point $(1,~-2,~-\frac{1}{2}(1+\sqrt{7}))$ :} Third point is a unstable critical point. Its effective EoS is the same as the model’s EoS.  Its unstable curves are similar to the first critical point.

{\bf \large Critical Point $(1,~-2,~\frac{1}{2}(\sqrt{7}-1))$ :} The final critical point is stable critical point. Its stable curves are already described in the second point.\\ 


\subsection{\large {\bf Model V :} $F(Q) =\eta Q^{2}$}
Here $\eta$ is a parameter \cite{Jim_nez_2020}. This model introduces late time cosmic acceleration without dark energy a and also gives rise to bouncing cosmologies (avoiding the Big Bang singularity). This model is noticed to modify the behavior of the early universe, possibly affecting inflation or the radiation/matter dominated eras \cite{gul2024comprehensive}.
As we know, in Friedmann-Leimatre-Robertson-Walker(FLRW) metric, the nonmetricity scalar $Q$ becomes
$$Q=6H^2~~\implies ~~f(Q)=36\eta H^4~~~~.$$
The standard Friedmann equation , hence, is altered to
$$3H^2=\rho_{eff}~~~~,$$
where the effective density $\rho_{eff}$ includes modified gravity theory contribution\cite{gadbail2024modified}. For this model,
\begin{equation}
F_Q (Q)=2 \eta Q=-\frac{y}{2}
\end{equation}

\begin{equation}
Q F_{QQ}(Q)=2 \eta Q=-\frac{y}{2}
\end{equation}

Again from the equations \eqref{x_derivative}, \eqref{y_derivative} and \eqref{u_derivative}, we get,

\begin{equation}
x^{\prime}=\frac{-3(x+y-1)(2x+y)}{-3y+2}~~~~~~,
\end{equation}

\begin{equation}
y^{\prime} = \frac{6 y (x+y-1)}{-3y+2}~~~~~~~~~and
\end{equation}

\begin{equation}
u^{\prime} = -u(u+2)- \frac{3(x+y-1)}{2-y}-\frac{3(x+y-1)u}{-3y+2}~~~~~~.
\end{equation}

Similarly, we solve $\begin{pmatrix}
{x}^\prime \\
{y}^\prime \\
{u}^\prime
\end{pmatrix}
= \tilde{0}$ and listed the critical points, eigen values of the matrix in table 5.
\begin{table}[h!] 
\caption{Critical points, Eigen Values of the Jacobian, equation of state parameter, expansion of matter disturbances and  Stability conditions for $F(Q) =\eta Q^{2}$.}
\centering
 \begin{tabular}{||c c c c c c c c||} 
 \hline
  Critical points & Eigen-values & $ \Omega_m$ & $ \Omega_Q$ & $ \omega_{eff}$ & $u$ & Equilibrium type & Stability condition. \\ [0.5ex] \hline
 $(1-y,~y,~-2)$ & $(0,~2,~-3)$ & $0$ & $1$ & $-1$ & $-2$ & non-hyperbolic & Saddle \\ 
 $(1-y,~y,~0)$  & $(-3,~-2,~0)$ & $0$ & $1$ & $-1$ & $0$& non-hyperbolic & Stable  \\
 $(0,~0,~-\frac{3}{2})$ & $(-3,~\frac{5}{2},~3)$ & $1$ & $0$ & $\omega$ & $\frac{-3}{2}$ & hyperbolic & Unstable \\
 $(0,~0,~1)$ & $(-3,~-\frac{5}{2},~3)$ & $1$ & $0$ & $\omega$ & $1$& hyperbolic & Unstable \\[1ex] 
 \hline
 \end{tabular}

\end{table}

\begin{figure}[h!]
$~~~~~~~~~~~~~~~~~~Fig-5(a)~~~~~~~~~~~~~~~~~~~~~~~~~~~~~~~~~~~Fig-5(b)~~~~~~~~~~~~~~~~~~~~~~~~~~~~~Fig-5(c)~~~$
\begin{center}
\includegraphics[scale=0.344]{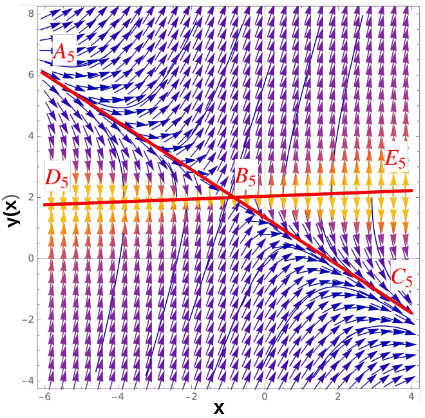}
\includegraphics[scale=0.375]{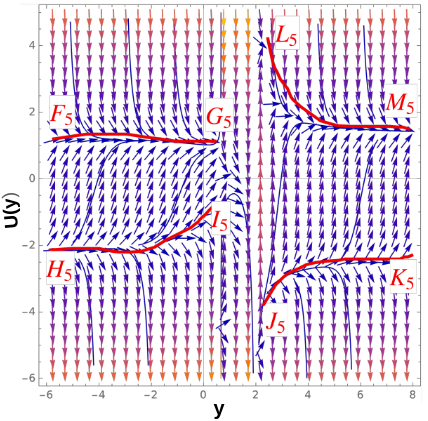}
\includegraphics[scale=0.36]{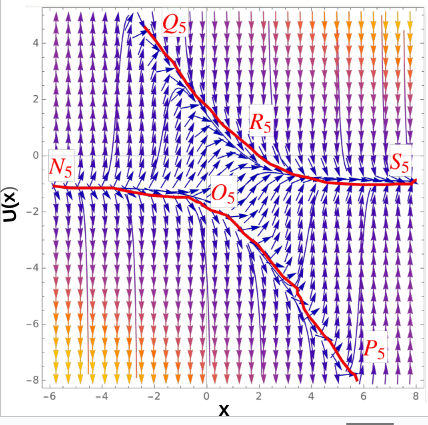}
\centering{Fig.$5(a)$-$5(c)$ : Phase portraits for the cosmological system obeying the non-metricity function $F(Q) =\eta Q^{2}$. Figure $5(a)$, $5(b)$ and $5(c)$ represent $y$ vs $x$, $u$ vs $y$ and $u$ vs $x$ spaces respectively.}
\end{center}
\end{figure}

{\bf \large Critical Point $(1-y,~y,~-2)$ :} This critical point is a saddle with cosmological constant like equation of state. Also, its matter contribution is null and the dark energy component takes the whole part of constituent energy density. $B_5$ is a saddle point in the phase portrait fig.$5(a)$. Matter evolution is also decaying at this point.

{\bf \large Critical Point $(1-y,~y,~0)$ :} This critical point also shows null matter contribution and effective EoS is the same as cosmological constant. This point is stable with two negative and one zero eigenvalues. Its evolution is constant. $B_5C_5$, $D_5B_5$, $F_5G_5$, $L_5M_5$ and $Q_5R_5S_5$ are the stable points in the phase portraits fig. $5(a)$-$5(c)$.

In this case, $x$ is the central variable and $y$, $u$ are the stable variables. The
corresponding matrices are $A = 0$ and $B= \begin{pmatrix}
-2 & 0\\  0 & -3
\end{pmatrix}$. The center manifold has now the form $y=h_1(x)$, and $u=h_2(x)$; the approximation $N$ has two components :

\begin{center}
$ N_1(h_1(x))=h_1^{\prime}(x)x^{\prime}-y^{\prime} = h_1^{\prime}(x) \frac{-3(x+y-1)(2x+y)}{-3y+2} - \frac{6 y (x+y-1)}{-3y+2}$\\

and $$ N_2(h_2(x))=h_2^{\prime}(x)x^{\prime}-u^{\prime} = h_2^{\prime}(x) \frac{-3(x+y-1)(2x+y)}{-3y+2} + u(u+2) + \frac{3(x+y-1)}{2-y}+\frac{3(x+y-1)u}{-3y+2}~~~.$$
\end{center}
For Zeroth approximation :
\begin{center}
$$ N_1(h_1(x)) = 0 +o(x^2)~~\text{and} ~~ N_2(h_2(x)) = \frac{3}{2}(x-1)+o(x^2)~~~.$$
\end{center}
The square model may represent a symmetric configuration (such as isotropy in the cosmic expansion) or even oscillatory behaviors in certain space time configurations. Again by examining the central manifold, we can capture the long-term behaviors, such as steady cosmic expansion, without detailed knowledge of each transient effect.

Non-hyperbolicity of these two fixed points are similar as those discussed before.

{\bf \large Critical Point $(0,~0,~-\frac{3}{2})$ :} This critical point refers to an unstable one. It has a null dark energy component. The matter component takes the whole part of constituent energy density. In the phase portraits fig. $5(a)$-$5(c)$ the unstable curves are $A_5B_5$, $B_5E_5$, $H_5I_5$, $J_5K_5$ and $N_5O_5P_5$. Since $u=-\frac{3}{2}$, thus the matter density $\delta$ varies as $a^{-\frac{3}{2}}$.

{\bf \large Critical Point $(0,~0,~1)$ :} This point has almost a similar interpretation except for the value of $u$. Since $u=-2$, thus the matter density $\delta$ varies as $a^{-2}$.\\ 

Let us discuss now regarding the differences of model $IV$ and $V$. As a cosmological model, $f(Q)=Q+\eta Q^2$ naturally replicates dark energy without cosmological constant. $f(Q)=Q^2$ requires a tuning to mimic late time acceleration. Depending on $\eta$, inflationary behavior is shown by model $IV$. Model $V$ generates inflation naturally. Model $V$ perturbates exotically while model $IV$ is more controlled. Fixed points of model $V$ is more unstable than those of $IV$. Both have nonhyperbolic fixed points independent of any parameters. Model $IV$ produces an unstable node as cosmological constant dominated universe. Model $V$ predicts the same epoch as a saddle.
\subsection{\large {\bf Model-VI :} $F(Q) =\alpha (-Q)^n$ }

Here $\alpha~~and~~n$ are two parameters \cite{Jim_nez_2020}. Plugging into this function, $$f(Q)=\alpha \left(-6H^2\right)^n~~~~.$$
Even though $Q=6H^2>0$, the negative sign is kept to make $(-Q)^n$ real for non integer $n$ and to support different conventions, where $R<0$ during expansion. A gravitational self energy is created that mimics dark energy. without any requirement of introduction of a seperate scalar field or cosmological constant \cite{solanki2022complete}. If $n=1$, GR is recovered without $\Lambda$. $0<n<1$ signifies $\omega_{eff}=\frac{p_{eff}}{\rho_{eff}}<-\frac{1}{3}$,i.e., quintessence and $n>1$ indicates either towards inflation or the phantom behavior \cite{sahlu2022linear}. Dynamical system for this model is formed as 
\begin{equation}
F_Q (Q)=-\alpha n (-Q)^{n-1} =-\frac{y}{2}~~~and
\end{equation}
\begin{equation}
Q F_{QQ}(Q)=-\alpha n(n-1)(-Q)^{n-1}=\frac{(1-n)y}{2}~~~.
\end{equation}
Again from the equations \eqref{x_derivative}, \eqref{y_derivative} and \eqref{u_derivative}, we get,
\begin{equation}
x^{\prime}=\frac{-3(x+y-1)(2x+y)}{(1-2n)y+2}~~~~~~~~,
\end{equation}
\begin{equation}
y^{\prime} = \frac{6 y (x+y-1)(n-1)}{(1-2n)y+2}~~~~~and
\end{equation}
\begin{equation}
u^{\prime} = -u(u+2)- \frac{3(x+y-1)}{2-y}-\frac{3(x+y-1)u}{(1-2n)y+2}~~~~~~~~~~~~~.
\end{equation}
Again solving $\begin{pmatrix}
{x}^\prime \\
{y}^\prime \\
{u}^\prime
\end{pmatrix}
= \tilde{0}$, we get the critical points which are listed in the table 6. For this model, depending on parameters, maximum of the equilibrium points fall in nonhyperbolic room.
\begin{table}[h!] 
\caption{Critical points, Eigen Values of the Jacobian, EoS parameter, expansion of matter disturbances and  Stability conditions for $F(Q) =\alpha (-Q)^n$.}
\centering
\begin{tabular}{||c c c c c c c c||} 
 \hline
  Critical points & Eigen-values & $ \Omega_m$ & $ \Omega_Q$ & $ \omega_{eff}$ & $u$ & Equilibrium type & Stability condition. \\ [0.5ex] \hline
 $(1-y,~y,~-2)$ & $(0,~2,~\frac{6-9y+6ny}{2+(1-2n)y})$ & $0$ & $1$ & $-1$ & $-2$ & non-hyperbolic & Unstable ($n=\frac{1}{2},y\leq 1$) $\&$\\
 & & & & & & & saddle ($n=\frac{1}{2},y>1$) \\ 
 $(1-y,~y,~0)$  & $(0,~-2,~\frac{6-9y+6ny}{2+(1-2n)y})$ & $0$ & $1$ & $-1$ & $0$  & non-hyperbolic & Saddle ($n=\frac{1}{2},y<1$) $\&$ \\
 & & & & & & & stable ($n=\frac{1}{2},y\geq 1$) \\
 ($0,~0,~-\frac{3}{2})$ & $(-3,~\frac{5}{2},~-3(-1+n))$ & $1$ & $0$ & $\omega$ & $\frac{-3}{2}$  & hyperbolic ($n\neq 1$) & Unstable for $n\neq 1$ $\&$ \\
   & & & & & & non-hyperbolic ($n= 1$) & saddle for $n= 1$\\[1ex] 
 $(0,~0,~1)$ & $(-3,~-\frac{5}{2},~-3(-1+n))$ & $1$ & $0$ & $\omega$ & $1$  & hyperbolic ($n\neq 1$) & Unstable for $n<1~\&$ \\
 & & & & & & & stable for $n> 1$ \\
  & & & & & & non-hyperbolic ($n= 1$) & stable for $n= 1$\\[1ex] 
 \hline
 \end{tabular}
\end{table}

\begin{figure}[h!]
$~~~~~~~~~~~~~~~~~~Fig-6(a)~~~~~~~~~~~~~~~~~~~~~~~~~~~~~~~~~~~~Fig-6b~~~~~~~~~~~~~~~~~~~~~~~~~~~~~~~~~~~~~~~~~~~~~~~~~~Fig-6(c)~~~$
\begin{center}
\includegraphics[scale=0.38]{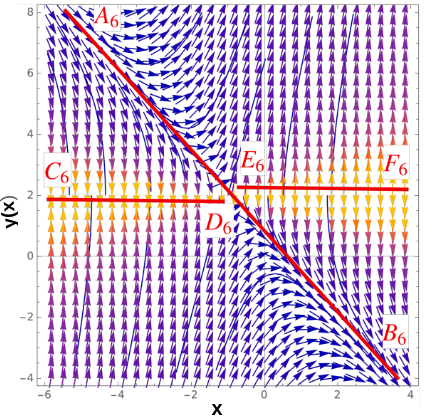}
\includegraphics[scale=0.36]{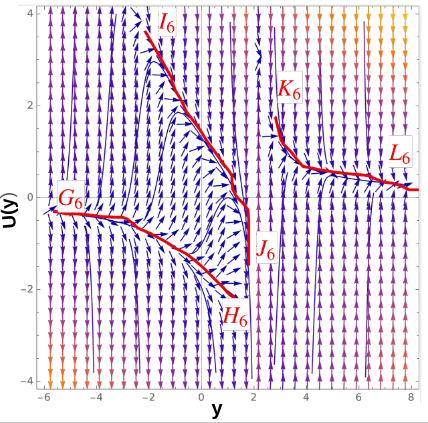}
\includegraphics[scale=0.35]{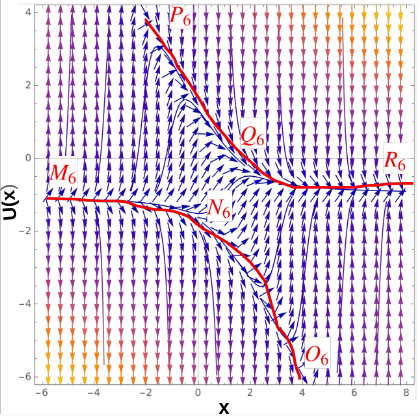}
\centering{Fig.$6(a)$-$6(c)$ : Phase portraits for the cosmological system obeying the non-metricity function $F(Q) =\alpha (-Q)^n$. Figure $(6a)$, $(6b)$ and $(6c)$ represent $y$ vs $x$, $u$ vs $y$ and $u$ vs $x$ spaces respectively.}
\end{center}
\end{figure}

{\bf \large Critical Point $(1-y,~y,~-2)$ :} The first critical point for $n=\frac{1}{2}$ is unstable for $y\leq 1$ and saddle for $y>1$. It also has null matter and whole dark energy contributions to the constituent energy density. Its EoS is also similar to the cosmological constant. Since $u=-2$, thus the matter density $\delta$ varies as $a^{-2}$. In the phase portraits fig. $6(a)$-$6(c)$, the unstable curves are $A_6E_6$, $E_6F_6$, $G_6H_6$ and $M_6N_6O_6$. 

{\bf \large Critical Point $(1-y,~y,~0)$ :} In the second case for $n=\frac{1}{2}$ the critical point is saddle for $y< 1$ and stable for $y\geq 1$. Its EoS is similar to the first critical point of model VI. The matter of evolution is constant. $E_6B_6$, $C_6D_6$, $I_6J_6$, $K_6L_6$ and $P_6Q_6R_6$ are the stable curves and $E_6$ is a saddle point.
 
{\bf \large Critical Point $(0,~0,~-\frac{3}{2})$ :} This critical point is the saddle for $n=1$ and unstable otherwise. The effective EoS is the same as the model’s EoS. Since $u=-\frac{3}{2}$, thus the matter density $\delta$ varies as $a^{-\frac{3}{2}}$.

{\bf \large Critical Point $(0,~0,~1)$ :} The final critical point is unstable for $n<1$ and stable for $n \ge 1$. The EoS is also the same as the model’s EoS. Since $u=1$, thus the matter density $\delta$ is directly proportional to $a$.\\ 

This model was the only among the lot studied in this article which has at least two equilibrium points which do change their nature depending on the value of $n$. This model is found to act as nonhyperbolic when $n\neq 1$, i.e., in the allowable range, when either quintessence, inflation or phantom era is chosen. GR, on the contrary, acts as a nonhyperbolic type for this particular model.

Let us suppose a quintessence model consists of a scalar field $\phi$ which comes with a suitable potential $V(\phi)=V_0exp\{-\lambda \phi\}$. An equilibrium point corresponding to the quintessence dominated epoch turns hyperbolic. Depending on the form of the potential or the dynamic variables' choice, many other cases may arise where quintessence dominated fixed points may turn hyperbolic, even stable \cite{Chakraborty2024}.

\section{Brief Discussion and Conclusion}
In this article, we have opted for a non-metricity scalar-dependent Einstein Hilbert action which generates a modified gravity theory. Such an $f(Q)$ dependent modified gravity has different faces depending on the type of $f(Q)$ function chosen. Different kinds of physical forms : $f(Q)=exp\{nQ\}~~,~~f(Q)=Q+\eta ln(\alpha Q)~~,~~f(Q)=Q+\eta Q^{-1}~~,~~f(Q)=Q+\eta Q^2~~\text{and}~~f(Q)=\eta Q^2~~,~~f(Q)=\alpha(-Q)^n$ are chosen to justify different cosmological scenario.

To understand the evolution underneath, we have constructed the related autonomous systems. Phase portraits of the related dynamical systems are drawn. To obtain the extremum points, the derivative of the column vector of dependent coordinates of the autonomous system is turned to vanish. These extremities are found to be spreaded through curves and sometimes at some points. As every point of a phase portrait signifies different states of the universe and the arrows are showing the tendency of their shift if a small perturbation is introduced to such points. In general, attractor and repeller lines and saddle points are found.

The first one, i.e., the exponential model which supports late time cosmic acceleration possesses hyperbolic saddle and stable equilibrium points which indicate towards cosmological phase transitions. These may occur at quark-hadron or electroweak transitions as at these critical events, equation of state of the universe changes. A non-hyperbolic equilibrium denotes a shift by small perturbations to lead a range of possible outcomes. Saddle may point towards the end of inflation which involves the shift of a dominant energy content. Hyperbolic attractor is found. These points require a constant Hubble parameter $H=\frac{\dot{a}}{a}=H_0$(say) and a deceleration parameter $q=-1$ which turns the scale factor $a(t)\propto exp\{H_0t\}$ which is a description of deSitter universe. The universe naturally evolves towards a stable accelerating phase. Matter dominated scaling solutions or Radiation dominated eras can also be justified under this case. Hyperbolic unstable points are significant in the study of $f(Q)$ gravity as they often serve as intermediate stages in the universe's evolution. For instance, the matter-dominated saddle point can transition to a stable dark energy dominated phase, reflecting the universe's shift from matter dominance to accelerated expansion.

In the second model, logarithmic term smoothens the contribution of $Q$. Like the previous case, hyperbolic and nonhyperbolic both type may occur.

Third model, comprises of an inverse of $Q$ term beside the GR equivalent linear term. Hyperbolic or nonhyperbolic, stable equilibrium points are not found for this universe.

Fourth model shows cosmological constant dominated universe as a nonhyperbolic one. Saddle may arise depending on parameters. Effective EoS for hyperbolic fixed points matches with that of dark energy. The transition to this regime from matter or radiation domination can involve such equilibrium. 
 
 Fifth and sixth models also exhibits different kinds of fixed points which are physically interpreted for cosmic evolution.

 A bifurcation in cosmology can signal a change from one type of universe to another. If we suppose a decelerating phase is transitting into an accelerating phase. Contraction to expansion may occur due to ``Big  Crunch" or ``Big bounce" etc in certain cyclic models. When such kind of bifurcations takes place near a nonhyperbolic equilibrium point, the end state sets so many options. Some among them act as stable attractors. This kind of regions, catalyses small perturbations to a large scale, qualitative changes in the evolution of the universe. The nonhyperbolic equilibrium indicates that the system is at a critical threshold from where the direction of evolution may change drastically.
\section{Acknowledgements}

Authors wish to thank IUCAA, Pune where the major part of this article was constructed. RB thanks IUCAA, Pune for Visiting Associateship.

\section{Data Availability Statement}
My manuscript has no associated data. [Authors’ comment: The authors have not used any data in the
manuscript. The authors have only compared the results with already openly available cosmological result.]
\newpage

\bibliographystyle{ieeetr} 

\bibliography{references}

\end{document}